\documentclass[copyright,creativecommons]{eptcs}
\usepackage{breakurl}        

\usepackage{amssymb}
\usepackage{amsfonts}
\usepackage{amsthm}
\usepackage{amsmath}
\usepackage{amstext}
\usepackage{enumitem}
\usepackage{latexsym}
\usepackage{stmaryrd}
\usepackage{makeidx}
\usepackage[all]{xy}

\newcommand{\Prime}[1]{\mathcal{P}\mathcal{R}(#1)}
\newcommand{\D}{{\mathcal D}}
\newcommand{\E}{{\mathcal E}}

\newtheorem{theorem}{Theorem}
\newtheorem{definition}[theorem]{Definition}

\newtheorem{lemma}[theorem]{Lemma}

\newtheorem{proposition}[theorem]{Proposition}

\newcommand{\cA}{\mathcal{A}}
\newcommand{\cB}{\mathcal{B}}
\newcommand{\cC}{\mathcal{C}}
\newcommand{\cD}{\mathcal{D}}
\newcommand{\cE}{\mathcal{E}}

\newcommand{\cP}{\mathcal{P}}

\newcommand{\ga}{\alpha}
\newcommand{\gb}{\beta}
\newcommand{\gc}{\gamma}
\newcommand{\gd}{\delta}

\newcommand{\gl}{\lambda}

\newcommand{\rf}{\mathrm{f}}

\newcommand\ev{\mathrm{ev}}
\newcommand\cur{\mathrm{cur}}
\newcommand{\id}{\textsf{id}} 

\newcommand{\dig}{\mathrm{dig}}
\newcommand{\codig}{\mathrm{codig}}
\newcommand{\der}{\mathrm{der}}
\newcommand{\coder}{\mathrm{cod}}

\newcommand{\Psd}{\mathbf{PSD}} 
\newcommand{\Sd}{\mathbf{SD}} 
\newcommand{\Palglat}{\mathbf{PAL}} 
\newcommand{\Cpo}{\mathbf{Cpo}} 
\newcommand{\Inf}{\mathbf{Inf}} 
\newcommand{\Inflfull}{\mathbf{InfLFull}} 
\newcommand{\Scottl}{\mathbf{ScottL}} 
\newcommand{\Rel}{\mathbf{Rel}} 
\newcommand{\Infl}{\mathbf{InfL}} 

\newcommand{\Con}{\mathrm{Con}} 

\title{On Linear Information Systems}

\author{
$\textrm{A.~Bucciarelli}^{\dagger}$
\institute{$\dagger$ Univerist\'{e} Paris Diderot \\ Paris, France}
\institute{Preuves Programmes et Syst\`{e}mes}
\email{Antonio.Bucciarelli@pps.jussieu.fr}
\and
$\textrm{A.~Carraro}^{\circ\dagger}$ \qquad\qquad
\institute{$\circ$ Univerist\`{a} Ca' Foscari \\ Venice, Italy}
\institute{Dipartimento di Informatica}
\email{acarraro@dsi.unive.it}
\and
$\textrm{T.~Ehrhard}^{\dagger}$ \qquad\qquad
\email{Thomas.Ehrhard@pps.jussieu.fr}
\and
$\textrm{A.~Salibra}^{\circ}$
\email{salibra@dsi.unive.it}
}

\def\titlerunning{On linear information systems}
\def\authorrunning{A. Bucciarelli \& A. Carraro \& T. Ehrhard \& A. Salibra}
\begin{document}
\maketitle

\begin{abstract}
Scott's information systems provide a categorically equivalent, intensional description of Scott domains and continuous functions. Following a well established pattern in denotational semantics, we define a linear version of information systems, providing a model of intuitionistic linear logic (a new-Seely category), with a ``set-theoretic" interpretation of exponentials that recovers  Scott continuous functions via the co-Kleisli construction. From a domain theoretic point of view, linear information systems are equivalent to prime algebraic Scott domains, which in turn generalize prime algebraic lattices, already known to provide a model of classical linear
logic. 
\end{abstract}

\section{Introduction}


The ccc of Scott domains and continuous functions, which we call $\Sd$, is the paradigmatic framework for denotational semantics of programming languages. In that area, much effort has been spent in studying more ``concrete" structures for representing domains. 

At the end of the 70's G. Kahn and G. D. Plotkin \cite{Kahn78} developed a theory of concrete domains together with a representation of them in terms of concrete data structures. In the early 80's G. Berry and P.-L. Curien \cite{Berry82} defined a ccc of concrete data structures and sequential algorithms on them. 
At the same time  Scott \cite{Scott82} also developed a representation theory for Scott domains which led him to the definition of information systems; these structures, together with the so-called approximable relations, form the ccc $\Inf$, which is equivalent to $\Sd$.

So it was clear that many categories of ``higher-level" structures such as domains had equivalent descriptions in terms of ``lower-level" structures, such as concrete data structures and information systems, which are collectively called webs.

At the end of the 80's, J.-Y. Girard \cite{Girard86} discovered linear logic starting from a semantical investigation of the second-order lambda calculus. His seminal work on the semantics of linear logic proofs \cite{Girard87, Girard88}, introduced a category of webs, the coherence spaces, equivalent to the category of coherent qualitative domains and stable maps between them. Coherence spaces form a $\ast$-autonomous and thus a model of classical Linear Logic (LL).

From the early 90's on, there has been a wealth of categorical models of linear logic, arising from different areas: we mention here S. Abramsky and R. Jagadeesan's games \cite{Abramsky92}, Curien's sequential data structures \cite{Curien94}, G. Winskel's event structures \cite{Winskel88, Zhang92}, and Winskel and Plotkin's bistructures \cite{Winskel94} whose associated co-Kleisli category is equivalent to a full-sub-ccc of Berry's category of bidomains \cite{Berry79}.

Remarkably, all the above-mentioned models lie outside Scott semantics. 

Despite the observation made by M. Barr's in 1979 \cite{Barr79} that the category of complete lattices and linear maps is $\ast$-autonomous, it was a common belief in the Linear Logic community that the standard Scott semantics could not provide models of classical LL, until 1994, when M. Huth showed \cite{Huth94} that the category $\Palglat$ of prime-algebraic complete lattices and lub-preserving maps is $\ast$-autonomous and its associated ccc $\Palglat^!$ (the co-Kleisli category of the ``!" comonad) is a full-sub-ccc of $\Cpo$. A few years later, Winskel rediscovered the same model in a semantical investigation of concurrency \cite{Winskel99,Winskel04}: indeed he showed that the category $\Scottl$ whose objects are preordered sets and the morphisms are functions from  downward closed to downward closed subsets which preserve arbitrary unions is $\ast$-autonomous; this category is equivalent to Huth's. T. Ehrhard \cite{Ehrhard09} continues this investigation and shows that the extensional collapse of the category $\Rel^!$, where $!$ is a comonad based on multi-sets over the category $\Rel$ of sets and relations, is the category $\Scottl^!$ and that both are new-Seely categories (in the sense of Bierman \cite{Bierman95}).

Summing up, there are several categorical models of LL in Scott semantics. In this paper we provide a representation of these models as webs. Our starting point are information systems, of which we provide a linear variant together with linear approximable relations: such data form the category $\Infl$, which we prove to be a symmetric monoidal closed category; $\Infl$ is equivalent to the category $\Psd$ of prime algebraic Scott domains and has as full-sub-categories $\Inflfull$ (equivalent to the category $\Palglat$) and, ultimately, $\Rel$.

We define a comonad $!$ over $\Infl$, based on sets rather than multi-sets, which makes $\Infl$ a new-Seely category and hence a model of intuitionistic MELL: our approach is different from that of \cite{Ehrhard09} in that our comonad is not an endofunctor of $\Rel$; we don't need to consider multisets exactly because we work in the bigger category $\Infl$. We also notice that $\Inflfull$ is the largest $\ast$-autonomous full-subcategory of $\Infl$ and that $\Infl^!$ is a full sub-ccc of $\Inf$.

\section{The category of linear informations systems}

Let $A$ be a set. We adopt the following conventions: letters $\ga, \gb, \gc, \ldots$ are used for elements of $A$; letters $a, b, c, \ldots$ are used for elements of $\cP_\rf(A)$; letters $x, y, z, \ldots$ are used for arbitrary elements of $\cP(A)$.

\begin{definition}
A \emph{linear information system} (LIS, for short) is a triple $\cA= (A, \Con, \vdash)$, where \mbox{$\Con \subseteq \cP_\rf(A)$} contains all singletons and $\vdash \ \subseteq A \times A$ satisfies the axioms listed below.
\begin{itemize}
\item[(IS1)] if $a \in \Con$ and $\forall \gb \in b.\exists \ga \in a.\ \ga \vdash \gb$, then $b \in \Con$
\item[(IS2)] $\ga \vdash \ga$
\item[(IS3)] if $\ga \vdash \gb \vdash \gc$, then $\ga \vdash \gc$
\end{itemize}
\end{definition}

The set $A$ is called the \emph{web} of $\cA$ and its elements are called \emph{tokens}. Let $\cA$, $\cB$ be two LISs. A relation $R \subseteq A \times B$ is \emph{linear approximable} if for all $\ga,\ga' \in A$ and all $\gb,\gb' \in B$:
\begin{itemize}
\item[(AR1)] if $a \in \Con_A$ and $\forall \gb \in b.\exists \ga \in a.\ (\ga, \gb) \in R$, then $b \in \Con_B$
\item[(AR2)] if $\ga' \vdash_A \ga \ R \ \gb \vdash_B \gb'$, then $(\ga', \gb') \in R$
\end{itemize}

Linear approximable relations compose as usual: $S \circ R = R; S = \{(\ga,\gc) \in A \times C : \exists \gb \in B.\ \ga \ R \ \gb \ S \ \gc\}$.

We call $\Infl$ the category with LISs as objects and linear approximable relations as morphisms. We reserve the name $\Inflfull$ for the full-subcategory of $\Infl$ whose objects are exactly those LISs $\cA$ for which $\Con_A = \cP_\rf(A)$. It is not difficult to see that the category $\Rel$ of sets and relations is a full-subcategory of $\Inflfull$, consisting of exactly those LISs $\cA$ for which $\Con_A = \cP_\rf(A)$ and \mbox{$\vdash_A = \{(\ga,\ga) : \ga \in A\}$}.

\subsection{The cartesian structure of $\Infl$}

We now define the cartesian product and coproduct in the category $\Infl$, which model the additive connectives of linear logic.

\medskip

\begin{definition}
Let $\cA_1$, $\cA_2$ be LISs. Define the LIS $\cA_1 \binampersand \cA_2 = (A_1 \binampersand A_2, \Con, \vdash)$ as follows:
\begin{itemize}
\item $A_1 \binampersand A_2 = A_1 \uplus A_2$
\item $\{(i_1,\gc_1), \ldots, (i_m, \gc_m)\} \in \Con$ iff $\{\gc_j : j \in [1,m],\ i_j = 1\} \in \Con_{A_1}$ and $\{\gc_j : j \in [1,m],\ i_j = 2\} \in \Con_{A_2}$
\item $(i, \gc) \vdash (j,\gc')$ iff $i = j$ and $\gc \vdash_{A_i} \gc'$
\end{itemize}
\end{definition}

The projections $\pi_i \in \Infl(\cA_1 \binampersand \cA_2, \cA_i)$ are given by \mbox{$\pi_i = \{((i,\gc), \gc') : \gc \vdash_{A_i} \gc' \}$}, $i = 1,2$. For \mbox{$R \in \Infl(\cC, \cA_1)$} and \mbox{$S \in \Infl(\cC, \cA_2)$}, the pairing \mbox{$\langle R, S \rangle \in \Infl(\cC, \cA_1 \binampersand \cA_2)$} is given by \\ \mbox{$\langle R, S \rangle = \{(\gc, (1,\ga)) : (\gc, \ga) \in R \} \cup \{(\gc, (2,\gb)) : (\gc, \gb) \in S \}$.} Define $\top = (\emptyset, \emptyset, \emptyset)$. The LIS $\top$ is the terminal object of $\Infl$, since for any LIS $\cA$, the only morphism $R \in \Infl(\cA, \top)$ is $\emptyset$.

\medskip

\begin{definition}
Let $\cA_1$, $\cA_2$ be LISs. Define the LIS $\cA_1 \oplus \cA_2 = (A_1 \oplus A_2, \Con, \vdash)$ as follows:
\begin{itemize}
\item $A_1 \oplus A_2 = A_1 \uplus A_2$
\item $\{(i_1,\gc_1), \ldots, (i_m, \gc_m)\} \in \Con$ iff $\{\gc_j : j \in [1,m],\ i_j = 1\} \in \Con_{A_1}$ and $\{\gc_j : j \in [1,m],\ i_j = 2\} \in \Con_{A_2}$
\item $(i, \gc) \vdash (j,\gc')$ iff $i = j$ and $\gc \vdash_{A_i} \gc'$
\end{itemize}
\end{definition}

Therefore cartesian products and coproducts coincide. The injections $\iota_i \in \Infl(\cA_i, \cA_1 \oplus \cA_2)$ are obtained reversing the projections, so that \mbox{$\iota_i = \{(\gc,(i,\gc')) : \gc \vdash_{A_i} \gc' \}$}, $i = 1,2$. For \mbox{$R \in \Infl(\cA_1, \cC)$} and \mbox{$S \in \Infl(\cA_2, \cC)$}, the ``co-pairing" \mbox{$[R, S] \in \Infl(\cA_1 \oplus \cA_2, \cC)$} is given by \\ \mbox{$[R, S] = \{((1,\ga), \gc) : (\ga, \gc) \in R \} \cup \{((2,\gb),\gc) : (\gb, \gc) \in S \}$.} Define $\mathbf{0} = \top = (\emptyset, \emptyset, \emptyset)$: this LIS is also the initial object of $\Infl$ and the unit of the coproduct.

\subsection{The monoidal closed structure of $\Infl$}

We now define the tensor product and its dual in the category $\Infl$, which model the multiplicative connectives of linear logic.

\medskip

\begin{definition}
Let $\cA$, $\cB$ be LISs. Define the LIS $\cA \otimes \cB = (A \otimes B, \Con, \vdash)$ as follows:
\begin{itemize}
\item $A \otimes B = A \times B$
\item $\{(\ga_1, \gb_1), \ldots, (\ga_m, \gb_m)\} \in \Con$ iff $\{\ga_1, \ldots, \ga_m\} \in \Con_A$ and $\{\gb_1, \ldots, \gb_m\} \in \Con_B$
\item $(\ga, \gb) \vdash (\ga',\gb')$ iff $\ga \vdash_A \ga'$ and $\gb \vdash_B \gb'$
\end{itemize}
\end{definition}

For $R \in \Infl(\cA, \cC)$ and $S \in \Infl(\cB, \cD)$, $R \otimes S \in \Infl(\cA \otimes \cB, \cC \otimes \cD)$ is given by \\ \mbox{$R \otimes S = \{((\ga,\gb), (\gc,\gd)) \in (A \otimes B) \times (C \otimes D) : (\ga,\gc) \in R \textrm{ and } (\gb,\gd) \in S \}$.} It is easy to check that \\ \mbox{$\_ \otimes \_ : \Infl \times \Infl \to \Infl$} is a bifunctor and that it is a symmetric tensor product, with natural isomorphisms
\begin{itemize}
\item $\phi_{\cA,\cB,\cC}^\otimes : \cA \otimes (\cB \otimes \cC) \to (\cA \otimes \cB) \otimes \cC$ given by 
$$ \phi_{\cA,\cB,\cC}^\otimes = \{( (\ga,(\gb,\gc)),((\ga',\gb'),\gc') ) : \ga \vdash_A \ga',\ \gb \vdash_B \gb',\ \gc \vdash_C \gc' \} $$

\item $\sigma_{\cA,\cB}^\otimes: \cA \otimes \cB \to \cB \otimes \cA$ given by $\sigma_{\cA,\cB}^\otimes = \{((\ga,\gb), (\gb',\ga')) : \ga \vdash_A \ga',\ \gb \vdash_B \gb'\}$

\item $\rho_\cA^\otimes: \cA \otimes \mathbf{1} \to \cA$ given by $ \rho_\cA^\otimes = \{((\ga, \ast),\ga') : \ga \vdash_A \ga'\}$

\item $\lambda_\cA^\otimes: \mathbf{1} \otimes \cA \to \cA$ given by $\lambda_\cA^\otimes = \{((\ast, \ga),\ga') : \ga \vdash_A \ga'\}$
\end{itemize}

Define $\mathbf{1} = (\{\ast\}, \{\emptyset, \{\ast\}\},\{(\ast, \ast)\})$. The LIS $\mathbf{1}$ is the unit of the tensor product.

The above data make $\Infl$ a symmetric monoidal category. As for the cartesian structure, also the dual of the tensor product, $\bindnasrepma$, coincides with the tensor in this category and thus the respective units $\bot$ and $\mathbf{1}$ are equal. So we take $\bot = (\{\ast\}, \{\emptyset, \{\ast\}\},\{(\ast, \ast)\})$ to be also the unit of $\bindnasrepma$.

\medskip

We now proceed to define the exponential objects of $\Infl$.

\begin{definition}
Let $\cA$, $\cB$ be LISs. Define the LIS $\cA \multimap \cB = (A \multimap B, \Con, \vdash)$ as follows:
\begin{itemize}
\item $A \multimap B = A \times B$
\item $\{(\ga_1, \gb_1), \ldots, (\ga_m, \gb_m)\} \in \Con$ iff for all $J \subseteq [1,m]$, $\{\ga_j : j \in J\} \in \Con_A$ implies $\{\gb_j : j \in J\} \in \Con_B$
\item $(\ga,\gb) \vdash (\ga',\gb')$ iff $\ga' \vdash_A \ga$ and $\gb \vdash_B \gb'$
\end{itemize}
\end{definition}

Define a natural isomorphism $\cur: \Infl(\cA \otimes \cC, \cB) \to \Infl(\cC, \cA \multimap \cB)$, (the linear currying) as $\cur(R) = \{(\gc, (\ga, \gb)) : ((\ga, \gc), \gb) \in R \}$. 
Define also the (linear) evaluation morphism \mbox{$\ev: \cA \otimes (\cA \multimap \cB) \to \cB$} as \mbox{$\ev = \{((\ga,(\ga',\gb)),\gb') : \ga \vdash_A \ga',\ \gb \vdash_B \gb' \}$}.

The above data make $\Infl$ a symmetric monoidal closed category, and thus a model of intuitionistic MLL proofs. The category $\Infl$ is however a rather degenerate model since the multiplicative connectives $\bindnasrepma$ and $\otimes$ coincide as well as the additives $\binampersand$ and $\oplus$.

\medskip

We now briefly discuss the issue of duality in the category $\Infl$. Let $\cA$ be a LIS and consider the LIS $\cA \multimap \bot$; an explicit description of such object is as follows: 
\begin{itemize}
\item $A \multimap \bot = A \times \{\ast\}$
\item $\Con_{A \multimap \bot} = \cP_\rf(A \times \{\ast\})$
\item $(\ga, \ast) \vdash_{A \multimap \bot} (\gb, \ast)$ iff $\gb \vdash_A \ga$
\end{itemize}

Therefore $\bot$ is not a dualizing object in $\Infl$, but it is so in $\Inflfull$, where the family of arrows $\partial_\cA: \cA \to (\cA \multimap \bot) \multimap \bot$ defined by $\partial_\cA = \{(\ga,((\ga',\ast),\ast)) : \ga \vdash_A \ga'\}$, for each LIS $\cA$, is a natural isomorphism. In other words $\Inflfull$ is the largest $\ast$-autonomous full-subcategory of $\Infl$.

\subsection{$\Infl$ is a new-Seely category}

In this section we define a comonad $!$ over $\Infl$ and prove that it gives a symmetric strong monoidal functor. Finally we prove that $\Infl$ is a new-Seely category and thus a model of intuitionistic MELL proofs, in which the exponential modality $!$ has a set-theoretic interpretation. For categorical notions we refer to Melli\`{e}s \cite{Mellies}.

\begin{definition}
Let $\cA$ be a LIS. Define the LIS $! \cA = (! A, \Con, \vdash)$ as follows:
\begin{itemize}
\item $! A = \Con_A$
\item $\{a_1, \ldots, a_k\} \in \Con$ iff $\cup_{i=1}^{k} a_i \in \Con_A$
\item $a \vdash b$ iff $\forall \gb \in b.\exists \ga \in a.\ \ga \vdash_A \gb$
\end{itemize}
\end{definition} 

Note that for $X \in !!A$ and $a \in !A$ we have $X \vdash_{!!A} \{b\}$ implies $\cup X \vdash_{!A} b$ but not viceversa. As an example, consider that $\{\alpha, \beta\} \vdash_{!A} \{\alpha, \beta\}$ but in general not $\{\{\alpha\}, \{\beta\}\} \vdash_{!!A} \{\{\alpha, \beta\}\}$.

Let $R \in \Infl(\cA, \cB)$. Define $! R \in \Infl(! \cA, ! \cB)$ as \mbox{$! R = \{(a, b) \in \ ! A \times ! B : \forall \gb \in b.\exists \ga \in a.\ (\ga, \gb) \in R\}$.} It is an easy matter to verify that \mbox{$!(\_): \Infl \to \Infl$} is a functor. Moreover ``$!$" is a comonad with digging \mbox{$\dig: ! \Rightarrow ! !$} defined by \mbox{$ \dig_\cA = \{(b, Y) \in \ ! A \times !! A : b \vdash_{!A} \cup Y \}$} and dereliction \mbox{$\der: ! \Rightarrow \textsf{id}_\Infl$} defined by \mbox{$\der_\cA = \{(b, \gb) \in \ !A \times A : b \vdash_{! A} \{\gb\} \}$.}

\medskip

As a matter of fact ``$!$" is also a monad if endowed with natural transformations \emph{codigging} \mbox{$\codig: !! \Rightarrow !$} defined by \mbox{$\codig_\cA = \{(X, a) \in \ !!A \times !A : X \vdash_{!!A} \{a\} \} = \der_{!\cA}$} and \emph{codereliction} \mbox{$\coder: \textsf{id}_\Infl \Rightarrow !$} defined by \mbox{$\coder_\cA = \{(\alpha, b) \in A \times !A : \{\alpha\} \vdash_{!A} b \}$.} This is due to the fact that \mbox{$a \vdash_{!A} b$} iff \mbox{$\{a\} \vdash_{!!A} \{b\}$} and \mbox{$X \vdash_{!!A} \{b\}$} iff \mbox{$\exists a \in X.\ a \vdash_{!A} b$}.

\medskip

This also shows a further symmetry: for each object $!\cA$, 
\begin{itemize}
\item the digging morphism is subsumed by the codereliction morphism in the sense that\\ \mbox{$\coder_{!\cA} = \{(a, X) : \{a\} \vdash_{!!A} X\} = \dig_\cA$}, since for $X \in !!A$ and $a \in !A$ we have \mbox{$\{a\} \vdash_{!!A} X$} iff \mbox{$a \vdash_{!A} \cup X$};
\item the codigging morphism is subsumed by the dereliction morphism in the sense that\\ \mbox{$\der_{!\cA} = \{(X, a) : X \vdash_{!!A} \{a\}\} = \codig_\cA$}.
\end{itemize}

\medskip

The forthcoming lemma shows that ``$!$" is a symmetric strong monoidal functor. Before proving this, we shall explicit the symmetric monoidal structure $(\Infl, \binampersand, \top)$ involved in the proof:
\begin{itemize}
\item $\phi_{\cA,\cB,\cC}^\binampersand : \cA \binampersand (\cB \binampersand \cC) \to (\cA \binampersand \cB) \binampersand \cC$ given by 
\begin{eqnarray*}
\phi_{\cA,\cB,\cC}^\binampersand & = & \big\{\big((1,\ga), (1,(1,\ga'))\big) : \ga \vdash_{A} \ga' \big\} \cup \big\{\big((2,(1,\gb)), (1,(2,\gb'))\big) : \gb \vdash_{B} \gb' \big\} \cup \\
 &  & \cup \big\{\big((2,(2,\gc)), (2,\gc')\big) : \gc \vdash_{C} \gc' \big\}
\end{eqnarray*}

\item $\sigma_{\cA,\cB}^\binampersand: \cA \binampersand \cB \to \cB \binampersand \cA$ given by 
$$ \sigma_{\cA,\cB}^\binampersand = \big\{\big((1,\ga), (2,\ga')\big) : \ga \vdash_{A} \ga' \big\} \cup \big\{\big((2,\gb), (1,\gb')\big) : \gb \vdash_{B} \gb' \big\} $$

\item $\rho_\cA^\binampersand: \cA \binampersand \top \to \cA$ given by $\rho_\cA^\binampersand = \big\{\big((1,\ga),\ga' \big) : \ga \vdash_A \ga' \big\}$

\item $\lambda_\cA^\binampersand: \top \binampersand \cA \to \cA$ given by $\lambda_\cA^\binampersand = \big\{\big((2,\ga),\ga' \big) : \ga \vdash_A \ga' \big\}$
\end{itemize}

\bigskip

\begin{lemma}\label{SSMF}
The functor $!: (\Infl, \binampersand, \top) \to (\Infl, \otimes, \mathbf{1})$ is symmetric strong monoidal.
\end{lemma}

\begin{proof}
We give the natural isomorphisms $\textsf{m}_{\cA,\cB}:\ ! \cA  \otimes ! \cB \cong !(\cA \binampersand \cB)$ and $\textsf{n}: \mathbf{1} \cong ! \top$ making $!$ a symmetric strong monoidal functor. Define 
\begin{itemize}
\item $\textsf{n} = \{(\ast, \{\emptyset\})\}$
\item $\textsf{m}_{\cA,\cB} = \big\{\big( (a,b), \{(1,\ga') : \ga' \in a' \} \cup \{(2,\gb') : \gb' \in b' \} \big) : a,a' \in \ ! A,\ b,b' \in \ ! B,\ a \vdash_{! A} a',\ b \vdash_{! B} b' \big\}$
\end{itemize}

We now proceed by verifying the commutation of the required diagrams. First observe that both $\textsf{m}_{\cA,\cB} \otimes \id_{! \cC} ; \textsf{m}_{\cA \binampersand \cB, \cC} ; ! \phi_{\cA,\cB,\cC}^\binampersand$ and $\phi_{\cA,\cB,\cC}^\otimes ; \textsf{id}_{! \cA} \otimes \textsf{m}_{\cB,\cC}  ; \textsf{m}_{\cA ,\cB \binampersand \cC}$ are equal to the set of all pairs 
$$\Big( ((a,b),c), \{(1,(1, \alpha_1)), \ldots, (1,(1, \alpha_{n_1})), (2,(1, \beta_1)), \ldots, (2, (1,\beta_{n_2})), (2,(2, \gamma_1)), \ldots, (2,(2, \gamma_{n_3}))\} \Big) $$
where $a \vdash_{!A} \{\alpha_1, \ldots, \alpha_{n_1}\}$, $b \vdash_{!B} \{\beta_1, \ldots, \beta_{n_2}\}$, and $c \vdash_{!C} \{\gamma_1, \ldots, \gamma_{n_3}\}$. Therefore the diagram

$$
\xymatrix{
(! \cA \otimes ! \cB) \otimes ! \cC \ar[r]^{\phi_{\cA, \cB, \cC}^\otimes} \ar[d]_{\textsf{m}_{\cA,\cB} \otimes \textsf{id}_{! \cC}} & ! \cA \otimes (! \cB \otimes ! \cC) \ar[d]^{\textsf{id}_{! \cA} \otimes \textsf{m}_{\cB,\cC}} \\
! (\cA \binampersand \cB) \otimes ! \cC \ar[d]_{\textsf{m}_{\cA \binampersand \cB,\cC}} & ! \cA \otimes ! (\cB \binampersand \cC) \ar[d]^{\textsf{m}_{\cA, \cB \binampersand \cC}} \\
! ((\cA \binampersand \cB) \binampersand \cC) \ar[r]^{! \phi_{\cA, \cB, \cC}^\binampersand} & ! (\cA \binampersand (\cB \binampersand \cC)) \\
}
$$
commutes. Finally the ``units" and the ``symmetry" diagrams
$$
\xymatrix{
! \cA \otimes \mathbf{1} \ar[r]^{\rho_{!\cA}^\otimes} \ar[d]_{\textsf{id}_{! \cA}  \otimes \textsf{n}} & ! \cA & & \mathbf{1} \otimes ! \cB \ar[r]^{\lambda_{!\cB}^\otimes} \ar[d]_{\textsf{n} \otimes \textsf{id}_{! \cB}} & ! \cB & & !\cA \otimes !\cB \ar[r]^{\sigma_{!\cA,!\cB}^\otimes} \ar[d]_{\textsf{m}_{\cA,\cB}} & !\cB \otimes !\cA \ar[d]^{\textsf{m}_{\cB,\cA}} \\
! \cA \otimes ! \top \ar[r]^{\textsf{m}_{\cA, \top}} & ! (\cA \binampersand \top) \ar[u]_{! \rho_\cA^\binampersand} & & ! \top \otimes ! \cB \ar[r]^{\textsf{m}_{\top, \cB}} & ! (\top \binampersand \cB) \ar[u]_{! \lambda_\cB^\binampersand} & & !(\cA \binampersand \cB) \ar[r]^{!\sigma_{\cA,\cB}^\binampersand} & ! (\cB \binampersand \cA)
}
$$
all commute because 
\begin{itemize}
\item $\id_{! \cA}  \otimes \textsf{n} ; \textsf{m}_{\cA, \top} ; !\rho_\cA^\binampersand = \{((a, \ast),a') : a \vdash_{! A} a'\} = \rho_{!\cA}^\otimes$,

\item $\textsf{n} \otimes \textsf{id}_{! \cB} ; \textsf{m}_{\top, \cB} ; !\lambda_\cB^\binampersand = \{((\ast, b),b') : b \vdash_{! B} b'\} = \lambda_{!\cB}^\otimes$,

\item both $\sigma_{!\cA,!\cB}^\otimes;\textsf{m}_{\cB,\cA}$ and $\textsf{m}_{\cA,\cB}; !\sigma_{\cA,\cB}^\binampersand$ equal the morphism \\ 
$\big\{\big( (a,b), \{(1,\gb') : \gb' \in b' \} \cup \{(2,\ga') : \ga' \in a' \} \big) : b' \in \ ! B,\ a' \in \ ! A,\ b \vdash_{! B} b',\ a \vdash_{! A} a' \big\}$.
\end{itemize}
\end{proof}

\begin{proposition}
$\Infl$ is a new-Seely category.
\end{proposition}

\begin{proof}
By Lemma \ref{SSMF}, $!$ is a symmetric strong monoidal functor; it remains to check the coherence diagram in the definition of new-Seely category. Indeed we have: 
\begin{eqnarray*}
\dig_\cA \otimes \dig_\cB ;  \textsf{m}_{! \cA,! \cB} & = & \big\{\big(  (a,b) ,  \{(1,a') : a' \in X'\} \cup \{(2,b') : b' \in Y'\} \big) : a \vdash_{! A} \cup X',\ b \vdash_{! B} \cup Y' \big\} \\
 & = & \textsf{m}_{\cA,\cB} ; \dig_{\cA \binampersand \cB}; ! \langle ! \pi_1, ! \pi_2 \rangle
\end{eqnarray*}
So that the following diagram commutes.
$$
\xymatrix{
! \cA \otimes ! \cB \ar[r]^{\textsf{m}_{\cA, \cB}} \ar[dd]_{\dig_\cA \otimes \dig_\cB} & ! (\cA \binampersand \cB) \ar[d]^{\dig_{\cA \binampersand \cB}} \\
 & !! (\cA \binampersand \cB) \ar[d]^{! \langle ! \pi_1, ! \pi_2 \rangle} \\
!! \cA \otimes !! \cB \ar[r]^{\textsf{m}_{!\cA, !\cB}} & ! (! \cA \binampersand !\cB)
}
$$
\end{proof}


\subsection{A representation theorem}

Not surprisingly, $\Infl$ turns out to be equivalent to the category of prime algebraic 
Scott domains and linear continuous functions. In this section, we outline this equivalence.

\begin{definition}
An element $p$ of a Scott domain $\cD$ is \emph{prime} if, whenever $B \subseteq_\rf D$ is upper bounded and $p \leq \vee B$, there exists $b \in B$ such that $p \leq b$. The domain $\cD$ itself is \emph{prime algebraic} if every element of $D$ is the least upper bound of the set of prime elements below it. The set of prime elements of $\cD$ is denoted by $\Prime{D}$.
\end{definition}

A \emph{linear function} between two prime algebraic Scott domains is a Scott continuos function that commutes with all (existing) least upper bounds. The category of prime algebraic Scott domains and linear function is denoted by $\Psd$.

\newpage

A \emph{point} of an information system $\cA$ is a subset $x \subseteq A$ satisfying the following two properties:
\begin{itemize}
\item[(PT1)] if $u \subseteq_\rf x$ then $u \in \Con$ ($x$ is \emph{finitely consistent})
\item[(PT2)] if $\alpha\in x$ and $\alpha \vdash \alpha'$, then $\alpha' \in x$ ($x$ is \emph{closed w.r.t.} $\vdash$)
\end{itemize}

We are now able to relate linear information systems to the corresponding categories of domains.

\begin{definition}\label{functors-for-equivalence}
Given $f,\cD,\cE, R ,\cA, \cB$ such that $f \in \Psd(\cD,\cE)$ and $R \in \Infl(\cA, \cB)$, we define:
\begin{itemize}
\item $\cA^+$ is the set of points of $\cA$ ordered by inclusion.

\item $R^+(x) = \{\beta \in B \mid \exists \alpha \in x.\ (\alpha,\beta)\in R\}$

\item $\cD^- = (\Prime{D},\Con,\vdash)$ where 
$a \in \Con$ iff $a$ is upper bounded and $p \vdash p'$ iff $p'\leq_\cD p$

\item $f^- = \{(p,p') \in \Prime{D} \times \Prime{E} \mid f(p) \geq_\cD p'\}$
\end{itemize}
\end{definition}

\begin{theorem}\label{equivalence}
The functors $(\_ )^+,(\_ )^-$ define an equivalence between the categories $\Infl$ and $\Psd$.
\end{theorem}

\begin{proof}
It is an easy task to check that $(\_ )^+: \Infl \to \Psd$ and ,$(\_ )^-: \Psd \to \Infl$ are indeed full and faithful functors and the two composite endofunctors $(-)^- \circ (-)^+$ and $(-)^+ \circ (-)^-$ are naturally isomorphic to the identity functor of $\Infl$ and $\Psd$, respectively. Moreover every hom-set $\Infl(\cA, \cB)$, ordered by inclusion of relations, is a prime algebraic Scott domain and $\Infl(\cA, \cB) = (\cA \multimap \cB)^+$, so that the functors $(-)^-$ and $(-)^+$ preserve exponentials, i.e. $(\cA \multimap \cB)^+ \cong \Psd(\cA^+, \cB^+)$ in the category $\Psd$ and $\Psd(\D, \E)^- \cong \D^- \multimap \E^-$ in the category $\Infl$. Finally the two functors also preserve products, since $(\cA \binampersand \cB)^+ \cong \cA^+ \times \cB^+$ in the category $\Psd$ and $(\D \times \E)^- \cong \D^- \binampersand \E^-$ in the category $\Infl$.
\end{proof}

In view of the categorical equivalence stated in Theorem \ref{equivalence}, Proposition \ref{co-Kleisli} shows how to recover Scott-continuous functions from linear ones. This equivalence specializes to an equivalence between $\Inflfull$ and the category $\Palglat$ of prime algebraic lattices and linear continuos functions.

\section{Classical versus linear information systems}

In the previous section we have treated a categorical equivalence explaining how linear information systems constitute a representation for prime algebraic Scott domains. Along the same lines Scott domains have an appealing representation as information systems introduced by Dana Scott in \cite{Scott82}. More recently \cite{Spreen08} a more general class of information systems have also been axiomatized, namely that of continuous information systems , defined in order to constitute a representation for continuous domains.

An information system consists of a set of tokens, over which are imposed an entailment and a consistency relation; it determines a Scott domain with elements those sets of tokens which are consistent and closed with respect to the entailment relation; the ordering is again just set inclusion. Vice versa a Scott domain defines an information system through its compact elements. In this section we review the basic notions on information systems.

Information systems are then organized in a category equivalent to that of Scott domains and continuous maps, $\Sd$, but more ``concrete'' and easier to work with under many respects. For example they have been used in \cite{Salibra09} to show that there is no reflexive object in $\Sd$ whose theory is exactly the least extensional lambda theory $\gl\gb\eta$.

In this section we explain the relation between the ``classical" Scott's information systems and linear information systems.

\newpage

Originally (\cite{Scott82}) an information system (IS, for short) is a triple $\cA= (A, \Con, \vdash)$, where \mbox{$\Con \subseteq \cP_\rf(A)$} contains all singletons and $\vdash \ \subseteq \Con \times \Con$ satisfies the axioms listed below.
\begin{itemize}
\item[(IS1)] if $a \in \Con$ and $a \vdash b$, then $b \in \Con$
\item[(IS2)] if $a' \subseteq a$, then $a \vdash a'$
\item[(IS3)] if $a \vdash b \vdash c$, then $a \vdash c$
\end{itemize}

\medskip

A relation $R \subseteq \Con_A \times \Con_B$ is \emph{approximable} if:
\begin{itemize}
\item[(AR1)] if $a \in \Con_A$ and $a \ R \ b$, then $b \in \Con_B$
\item[(AR2)] if $a' \vdash_A a \ R \ b \vdash_B b'$, then $a'\ R\ b'$
\end{itemize}

Clearly the every approximable relation, included $\vdash$, is completely determined by tokens on the right-hand side in the sense that $a \ R \ b$ iff $\forall \beta \in b.\ a \ R \ \{\beta\}$. Hence we shall identify each approximable relation $R$ with its \emph{trace} $\{(a, \beta) : (a, \{\beta\}) \in R \}$.

\medskip

We call $\Inf$ the category with ISs as objects and approximable relations as morphisms. It is well-known that $\Inf(\cA, \cB)$, ordered by inclusion of relations, is a Scott domain. Let us recall the definition of exponentials in $\Inf$ (\cite{Larsen91}).

\begin{definition}
Let $\cA$, $\cB$ be ISs. Define the IS $\cA \Rightarrow \cB = (A \Rightarrow B, \Con, \vdash)$ as follows:
\begin{itemize}
\item $A \Rightarrow B = \Con_A \times B$
\item $\{(a_1, \beta_1), \ldots, (a_m, \beta_m)\} \in \Con$ iff for all $J \subseteq [1,m]$, $\cup \{a_j : j \in J\} \in \Con_A$ implies $\{\beta_j : j \in J\} \in \Con_B$
\item $\{(a_1, \beta_1), \ldots, (a_m,\beta_m)\} \vdash (a', \beta')$ iff $\{\beta_j : a' \vdash a_j,\ 1 \leq j \leq m \} \vdash_B \beta'$
\end{itemize}
\end{definition}

Similarly to the functors given in Definition \ref{functors-for-equivalence}, there are functors $(\_ )^\bullet: \Inf \to \Sd$ and \mbox{$(\_ )^\circ: \Sd \to \Inf$} which define another equivalence of categories. In particular for a given IS $\cA$, $\cA^\bullet$ is the collection of all subsets $x \subseteq A$ satisfying the following two properties:
\begin{itemize}
\item[(PT1')] if $u \subseteq_\rf x$ then $u \in \Con$ ($x$ is \emph{finitely consistent})
\item[(PT2')] if $ a \subseteq x$ and $a \vdash \alpha'$, then $\alpha' \in x$ ($x$ is \emph{closed w.r.t.} $\vdash$)
\end{itemize}

Again we have that $\Inf(\cA, \cB) = (\cA \Rightarrow \cB)^\bullet$. The categorical equivalence stated in Theorem \ref{equivalence} mirrors perfectly the equivalence between the categories $\Inf$ and $\Sd$, so that the definition of linear information system and linear approximable relation is exactly what is required in order to capture the passage from the category of Scott domains and continuous functions to that of prime algebraic Scott domains and linear functions.

In fact both linear information systems and linear approximable relations can be seen as particular information systems and approximable relations, respectively, exactly as prime algebraic Scott domains are Scott domains and linear functions are continuous functions. The next proposition, based on this fact, shows that again, following a well-established pattern, the comonad $!$ allows to recover non-linear approximable relations form linear ones. As usual, we denote by $\Infl^!$ the co-Kleisli category of the comonad $!$ over $\Infl$.

\begin{proposition}\label{co-Kleisli}
$\Infl^!$ is a full-sub-ccc of $\Inf$.
\end{proposition}

\begin{proof}
Let $\cA,\cB$ be LISs and let $\cA \Rightarrow \cB$ be the exponential object formed in the category $\Inf$: $\cA \Rightarrow \cB$ is a linear information system and it is an easy matter to see that $\cA \Rightarrow \cB = !\cA \multimap \cB$. Moreover $\cC^+ = \cC^\bullet$, for any LIS $\cC$, and thus \mbox{$\Infl(!\cA, \cB) = (!\cA \multimap \cB)^+ = (\cA \Rightarrow \cB)^\bullet = \Inf(\cA, \cB)$}.
\end{proof}

The space $\Infl(\cA, \cB)$ of clearly embeds into $\Inf(\cA, \cB)$ (exactly as the space of linear functions embeds into that of continuous functions). The embedding is given by the map $\varphi: \Infl(\cA, \cB) \hookrightarrow \Inf(\cA, \cB)$ given by $\varphi(R) = \{(a, \beta) : \exists \alpha \in a.\ \alpha \ R \ \beta \}$. In other words the linear approximable relations are elements of $\Inf(\cA, \cB)$, i.e. exactly those approximable relations $S$ for which $(a, \beta) \in S$ iff $\exists \alpha \in a.\ (\alpha, \beta) \in S$. This is the analogue of the condition, dealing with preservation of existing suprema, that isolates linear functions between Scott domains among the continuous ones. 

\section{Conclusions and future work}

In this paper we defined the category $\Infl$, whose objects and arrows result from a linearization of Scott's information systems. 
We show this category to be symmetric monoidal closed and thus a model of MLL. We moreover prove that $\Infl$ is a new-Seely category, 
with a ``set-theoretic'' interpretation of exponentials via a comonad $!$; this is made possible by the presence of the entailment relation, 
which is always non-trivial in objects of the form $!\cA$:
even if the entailment $\vdash_A$ is the equality we have that $\Infl(!\cA, \cA) \subset \cP(!\cA \times \cA)$ and this rules out Ehrhard's 
counterexample for the naturality of dereliction. In the purely relational model of classical MELL, $\Rel$, 
the use of multisets is needed. 

Indeed a comonad based on multi-sets, let's say $\dag$, can be defined in our framework too, yelding a different co-Kleisli category.
A similar situation arises in the framework of the coherence spaces model of LL. In that case Barreiro and Ehrhard \cite{Barreiro97} 
proved that the extensional collapse of the hierarchy of simple types associated to the multi-set interpretation {\em is} the hierarchy associated to the ``set'' interpretation. This means in particular that, as models of the simply typed $ \lambda$-calculus, the former discriminates finerly than the latter,
being sensitive, for instance, to the number of occurrences of a variable in a term. 
It is likely that, in a similar way, $\Infl^!$ is the extensional collapse of $\Infl^\dag$.

The categories $\Infl$ and $\Inflfull$ may be themselves compared using the same paradigm. Trivializing the consitency relation boils down to add points to the underlying domains; is that another instance of extensional collapse situation? In the case of  the simple types hierarchy  over the booleans, the Scott model is actually the extensional collapse of the lattice-theoretic one \cite{Bucciarelli96}.

Summing up it appears that, by  tuning the linear information systems in different ways, one  obtains different frameworks for the interpretation of proofs, whose inter-connections remain to be investigated.

This is, in our opinion, the main advantage of the approach, with respect to the existing descriptions of the Scott continuous models of linear logic \cite{Barr79, Huth94, Huth94b, Winskel99}. 

Linear information systems are close to several classes of webbed models of the pure the $\lambda$-calculus: they generalize Berline's {\em preordered sets with coherence} \cite{Berline00}, where a set of tokens is consistent if and only if its elements are pairwise coherent. We plan to investigate whether such a generalisation is useful for studying the models of the $\lambda$-calulus in  $\Psd$. Actually, one of our motivations was to settle a representation theory for a larger class of cartesian closed categories, whith $\Rel^!$ as a particular case, in order to provide a tool for investigating ``non-standard''\footnote{Let us call ``standard'' a  model of the $\lambda$-calculus that is an instance of one of the ``main'' semantics: continuous, stable, strongly stable.} models. We get $\Rel$ as a full subcategory of $\Infl$, but the bunch of axioms on information-like structures making $\Rel^!$ an instance of the co-Kleisli construction remains to be found.

Another original motivation for investigating linear information systems, that we are pursuing, was the definition of a framework suitable for the interpretation of Boudol's $\lambda$-calculus with resources \cite{Boudol93}.

Finally we point out another research direction, suggested by the work of Ehrhard and Regnier (\cite{Ehrhard06}, for example). In K\"{o}the spaces \cite{Ehrhard02} as well as in finiteness spaces \cite{Ehrhard05}, linear logic formulae are interpreted as topological vector spaces, and proofs of linear logic as linear continuous maps between these spaces. Then exponentials appear as ``symmetric tensor algebra" constructions \cite{Blute93}. In the models considered there, linear maps from $!X$ to $Y$ can be seen as ``analytic functions" (that is, functions definable by a power series) from the vector space $X$ to the vector space $Y$ and therefore can be differentiated.  Classically, the derivative of a function $f: X \to Y$ is a function $f: X \to (X \multimap Y)$ such that for each $x \in X$, the linear function $f'(x)$ (the derivative of $f$ at point $x$) is the ``best linear approximation" of the function $X \to Y$ which maps $u \in X$ to $f(x + u) \in Y$ (the general definition is local). In the analytic case, differentiation turns a linear function $f: !X \to Y$ into a linear function $f': !X \to (X \multimap Y)$, that is, $f': (!X \otimes X) \to Y$. It turns out that $f'$ can be obtained from $f$ by composing it (as a linear function from $!X$ to $Y$) on the left with a particular linear morphism $d: (!X \otimes X) \to !X$. This morphism itself can be defined in terms of more primitive operations on $!X$. 

This can certainly be done also in the category $\Infl$ following for \cite{Hyland03}, for example. However the induced differential combinator in this case does not reflect the idea of approximation typical of Scott semantics. The purpose we have in mind is to investigate the possibility of symmetric tensor bialgebra constructions, in the category $\Infl$, giving rise in the equivalent category $\Psd$ to a reasonable notion of derivative and compatible with the usual idea of approximation in Scott semantics.



\bibliographystyle{eptcs}

\end{document}